\documentclass[final,5p,twocolumn,times]{elsarticle}
\usepackage{amsmath,amssymb,bm,color,graphicx,hyperref} 
\usepackage{flushend}

\DeclareMathOperator{\tr}{tr}

\DeclareMathOperator{\dd}{d\!}
\DeclareMathOperator{\sgn}{sgn}

\newcommand{\ri}{\mathrm{i}}

\newcommand{\re}{\mathrm{e}}

\newcommand{\bsf}[1]{\bm{\mathsf{#1}}}

\journal{Mechanics Research Communications}

\begin{document}

\begin{frontmatter}

\title{Stress retardation versus stress relaxation in linear viscoelasticity}

\author[icc]{Ivan C. Christov\corref{cor1}}
\ead{christov@purdue.edu}
\ead[url]{http://christov.tmnt-lab.org}
\address[icc]{School of Mechanical Engineering, Purdue University, West Lafayette, IN 47907, USA}

\author[cic]{C. I. Christov\fnref{fn1}}
\ead[url]{http://christov.metacontinuum.com}
\address[cic]{Department of Mathematics, University of Louisiana at Lafayette, Lafayette, LA 70504, USA}

\fntext[fn1]{Prof.~C.~I.~Christov passed away prior to submission of the manuscript. This paper is dedicated to his memory.}
\cortext[cor1]{To whom correspondence should be addressed.}


\begin{abstract}
We present a preliminary examination of a new approach to a long-standing problem in non-Newtonian fluid mechanics. First, we summarize how a general implicit functional relation between stress and rate of strain of a continuum with memory is reduced to the well-known linear differential constitutive relations that account for ``relaxation'' and ``retardation.'' Then, we show that relaxation and retardation are asymptotically equivalent for small Deborah numbers, whence causal pure relaxation models necessarily correspond to ill-posed pure retardation models. We suggest that this dichotomy could be a possible way to reconcile the discrepancy between the theory of and certain experiments on viscoelastic liquids that are conjectured to exhibit only stress retardation. 
\end{abstract}


\end{frontmatter}

\section{Introduction} Viscoelastic non-Newtonian fluids continue to be an active area of research not only because of the difficulties in their theoretical modeling \cite{Petrie} and the challenges in their experimental interrogation \cite{KhairSquires}, but also because of their abundance in biophysics \cite{Teran,PRE_squirm,Spag15} and their relevance to continua with local thermal non-equilibrium effects \cite[\S8.4]{S15_LTNE}.

Recently, new experimental methods have been proposed for rheological measurements of polymeric solutions \cite{KhairSquires} and novel calculations have been performed for the locomotion of microorganisms in ``weakly viscoelastic'' fluids \cite{PRE_squirm}. Yet, the ``second-order fluid'' model used in the latter works, and also for interpreting previous experiments \cite{MarkovitzBrown}, is unstable (ill-posed in the sense of Hadamard) \cite{ColemanDuffinMizel,DunnFos,Joseph} for a first normal stress difference $\Psi_1>0$ as measured. Various explanations have been put forth \cite{DunRaj}, often questioning the experimental setup and data analysis. Others dismiss the difficulty as not important for ``small'' departures from Newtonian behavior. Similar ill-posed models arise from the Chapman--Enskog expansion of the Boltzmann--Bhatnagar--Gross--Krook equation when keeping only leading-order non-Newtonian terms \cite{Yakhot,Yakhot_PRL}.

In the face of such extensive evidence that, in the real world, the first normal stress difference $\Psi_1>0$ for a second-order fluid, it appears to us that it is neither satisfactory to claim that the instability is not manifested for ``slow flows'' or ``small departures from Newtonian behavior'' nor is it satisfactory to repeat the mantra that all experimental results are inconclusive or wrong. New insights are needed to understand such a non-trivial discrepancy in the foundations of viscoelasticity, given the resurgence of the ``second-order fluid'' model \cite{KhairSquires,PRE_squirm,Yakhot,Yakhot_PRL}. In this preliminary research report, we propose another approach. Specifically, we show how the ill-posed second-order (retardational) fluid model may arise as an improper interpretation of a fluid that is actually exhibiting stress relaxation of the Maxwell type \cite{M67},\footnote{Maxwell-type relaxation is also common in nonclassical theories of heat conduction \cite{JosephPrez,JosephPrezAdd,S11_HW} and thermoelasticity \cite{IOS10}.} since the latter would be indistinguishable from the former for small departures from Newtonian rheology. 

\section{Background on memory effects and nonlocal rheology} In this section, in order to make this preliminary research report self-contained and accessible to a wider audience, we summarize the standard background on constitutive modeling for viscoelastic fluids.

As usual, we decompose the stress tensor $\bsf{T}$ into an indeterminate part (the spherical pressure $p$) and a constitutive part $\bsf{S}$ as $\bsf{T} = -p\bsf{I} + \bsf{S}$. We consider only isochoric motions (or incompressible fluids) so that $\tr(\bm{\nabla}\bm{u}) = \bm{\nabla}\bm{\cdot}\bm{u} = 0$, where $\bm{u}$ is the velocity field. The fluid is assumed homogeneous and isotropic so that it has constant density $\varrho_0$, and its rheological parameters (e.g., the viscosity) are constant scalars.

The most general implicit relationship between the stress tensor $\bsf{T}(\bm{x},t)$ and the rate-of-strain tensor $\bsf{E}(\bm{x},t)$ that includes the effect of \emph{memory} is a \emph{functional} that depends on the independent variables.
The relationship is further assumed to be local in the spatial variable (i.e., the functional's value at a given point $\bm{x}$ is a point function of these tensors at $\bm{x}$) to preclude ``action at a distance'' effects. Hence, 
\begin{equation}
\mathfrak{F}[\bsf{S}(\bm{x},\cdot),\bsf{E}(\bm{x},\cdot)](\bm{x},t) = const.,
\label{eq:implicit}
\end{equation} 
where $\mathfrak{F}$ is a continuous functional, and the ``dummy'' variable of integration is substituted in place of the dots. 

Equation~\eqref{eq:implicit} can be developed into a \emph{Volterra functional series} (see, e.g., Walters~\cite{Walters} and Bird et al.~\cite[\S9.6]{Bird}):
\begin{multline}
M^{(0)}(t)  + \int\limits_{-\infty}^t M^{(1)}(t-s;t) \bsf{S}(\bm{x},s) \dd s + \cdots\\ + \sum_{j=2}^\infty \idotsint\limits_{-\infty}^{\quad t} \frac{1}{j!}M^{(j)}(t-s_1,\dots,t-s_j;t)  \prod_{l=1}^j \bsf{S}(\bm{x},s_l)\dd s_l \\
 = K^{(0)}(t) + \int\limits_{-\infty}^t K^{(1)}(t-s;t) \bsf{E}(\bm{x},s) \dd s + \cdots\\ +\sum_{j=2}^\infty \idotsint\limits_{-\infty}^{\quad t} \frac{1}{j!}K^{(j)}(t-s_1,\dots,t-s_j;t) \prod_{l=1}^j \bsf{E}(\bm{x},s_l)\dd s_l.
\label{eq:general}
\end{multline}
Let us further assume that the constitutive relation \eqref{eq:implicit} does not depend explicitly on time, i.e., the functional $\mathfrak{F}$ is \emph{stationary}, or \emph{time invariant} \cite{BoydChuaDesoer}, so that the kernels $M^{(0)},K^{(0)} = const.$, and the kernels $M^{(j)},K^{(j)}$ are functions of the ``dummy'' variable only. Since the fluid is isotropic, the kernels are scalar functions of their argument.\footnote{The kernels $M^{(j)}$ and  $K^{(j)}$ are related to the Fr\'echet derivatives of the functional $\mathfrak{F}$ in \eqref{eq:implicit} \cite{BoydChuaDesoer}, which makes the Volterra expansion analogous to a Taylor series. Its convergence is beyond the scope of the present work, however.
} Also, requiring that zero stress produces zero strain (i.e., we do not consider plasticity), together with the time-invariance of the constitutive relation, implies that  $M^{(0)} = K^{(0)} = 0$. 

Equation~\eqref{eq:general} is the most general nonlocal (functional) dependence of the rate of stress on the strain as first proposed by Green and Rivlin~\cite{GreenRivlin57} from a different perspective. The memory effects are modeled for all time, i.e., from $t=-\infty$, without loss of generality, since a cut-off from fading (or somehow limited) memory can be introduced through the kernels. The upper limit of integration is $t$ so that the relation is causal, i.e., $\bsf{S}$ (and therefore $\bsf{T}$) depends only on the values of $\bsf{E}$ for the instants of time prior to the current one.

\subsection{Linearized memory relations} When the functional $\mathfrak{F}$ in \eqref{eq:implicit} is \emph{linear} in its two arguments, \eqref{eq:general} reduces to
\begin{equation}
 \int_{0}^\infty M(\zeta) \bsf{S}(t-\zeta) \dd \zeta = \int_0^{\infty} K(\zeta) \bsf{E}(t-\zeta) \dd \zeta
 \label{eq:linear_implicit}
\end{equation}
after the change of variables $\zeta = t-s$. The superscript ``(1)'' on the kernels is omitted for the sake of simplicity of notation. Furthermore, for consistency with Navier--Stokes theory, we assume that $\int_{0}^\infty M(\zeta) \dd \zeta = 1$ and $\int_0^\infty K(\zeta) \dd \zeta \ne 0$. Under mild restrictions on the kernels, one can resolve \eqref{eq:linear_implicit}, using the Laplace transform and the convolution theorem, into $\bsf{S} =\int_0^\infty \mathcal{K}(\zeta) \bsf{E}(t-\zeta) \dd \zeta$ (strain memory only) or $\bsf{E} = \int_0^\infty \mathcal{M}(\zeta) \bsf{S}(t-\zeta) \dd \zeta$ (stress memory only).
The former case is related to the classic memory assumption of Coleman and Noll~\cite{ColemanNoll,ColemanNollErr}, which is recovered if a Dirac delta is stipulated to be part of the resolved kernel. The latter case gives the implicit ``twin'' of the Coleman--Noll theory. Though the kernels $M$ and $K$ in \eqref{eq:linear_implicit} may be well-behaved for fast fading memory, after the resolution with respect to either $\bsf{S}$ or $\bsf{E}$, the effective kernels $\mathcal{M}$ and $\mathcal{K}$ do not necessarily have the same smoothness properties. In other words, \emph{it may not always be desirable to separate relaxation from retardation in the general linear constitutive relation \eqref{eq:linear_implicit}.}

\subsection{Differential constitutive relations} Constitutive relations involving derivatives of $\bsf{S}$ and $\bsf{E}$ have been used extensively in the last couple of decades \cite{Bird_ARFM}. To motivate such differential approximations of the rheology with memory, we expand the tensors $\bsf{S}(t-\zeta)$ and $\bsf{E}(t-\zeta)$ into Taylor series about $t=0$ (see also \cite{Christov_BeyondFourier} for a related derivation in the hyperbolic heat conduction context):
\begin{equation}
\bsf{S}(t-\zeta) = \sum_{j=0}^\infty \frac{(-\zeta)^j}{j!} \bsf{S}^{(j)}(t), \quad \bsf{E}(t-\zeta) = \sum_{j=0}^\infty \frac{(-\zeta)^j}{j!} \bsf{E}^{(j)}(t).
\end{equation}
Substituting the latter expressions into \eqref{eq:linear_implicit}, we obtain
\begin{equation}
\bsf{S} + \tau_1 \dot{\bsf{S}} + \tau_2 \ddot{\bsf{S}} + \cdots = \mu_0\big(\bsf{E} + \mu_1 \dot{\bsf{E}} + \mu_2 \ddot{\bsf{E}} + \cdots\big),
\label{eq:differential}
\end{equation}
where $\tau_0=1$, $\tau_j := \frac{(-1)^j}{j!} \int_{0}^{\infty} \zeta^j {M}(\zeta) \dd \zeta$ ($j\ge1$), $\mu_0 = \int_0^\infty K(\zeta) \dd \zeta$ and $\mu_0\mu_j := \frac{(-1)^j}{j!} \int_{0}^{\infty} \zeta^j {K}(\zeta) \dd \zeta$ ($j\ge1$); $\tau_j,\mu_j$ ($j\ge 1$) carry units of time$^j$, while $\mu_0(>0)$ is the viscosity understood in the sense of Navier--Stokes theory. 
The general differential constitutive relation \eqref{eq:differential} was anticipated by Burgers \cite{Burgers}.

The terms with derivatives on the left-hand side of \eqref{eq:differential} are called (``generalized'') \emph{relaxations}, while the respective terms on the right-hand side of \eqref{eq:differential} are termed (``generalized'') \emph{retardations}.\footnote{Another name for the physical effect described by the word `retardation' is \emph{elastic hysteresis} due to internal friction \cite[p.~19]{Burgers}.} Respectively, the coefficients $\tau_j$ are the ``generalized relaxation times,'' while the $\mu_j$ are the ``generalized retardation times.'' Note that we have changed the primes to dots in order to emphasize the fact that these are derivatives with respect to $t$. For the present purposes, it suffices to identify these with ordinary time derivatives, and henceforth $\dot{(\cdot)} \equiv \partial_t(\cdot) \equiv \partial(\cdot)/\partial t$. However, going beyond unidirectional flows in stationary media, one has to replace them with properly invariant convected time rates \cite{Oldroyd,Truesdell,CIC_MRC,OS09}. 

Finally, it is important to note that a nonlocal rheology of differential type may only be used when all the integrals defining each $\tau_j$ and $\mu_j$ exist. The issue was brought up by Coleman and Markovitz~\cite[\S 2]{ColemanMarkovitz} and elucidated further by Joseph~\cite{Joseph}. This means that the decay of the kernel at infinity must be super-algebraic (unless the expansion is truncated at some finite $j$); the simplest case is that of exponential decay \cite{Coleman_waves1,Coleman_waves2,Coleman_waves3,Coleman_waves4}.\footnote{If the fading memory follows a power law $\zeta^{-\beta}$, $\beta\in(0,1)$, then even the integral defining $\tau_1$ and/or $\mu_1$ can diverge, and the differential constitutive relation will feature a fractional-order derivative, if it exists at all. In heat conduction through a polydisperse suspension (see, e.g., \cite{ChowChri_PRSA}), one has $\frac{1}{\Gamma(\beta)}\int_{0}^{t} (t-s)^{\beta-1}\bsf{E}(s) \dd s \equiv {}_0D_t^{-\beta} \bsf{E}$, i.e., the Riemann--Liouville fractional integral \citep[\S1.1]{Maina}, as the right-hand side of \eqref{eq:linear_implicit}.} In this case, the differential approximation can be especially good quantitatively since only the first few coefficients $\mu_j$ are non-negligible, i.e., $\mu_j \propto \epsilon^{j+1}$ for a kernel $\propto\re^{-\zeta/\epsilon}$ with $\epsilon$ ``small.'' 

\section{Asymptotic equivalence of relaxation and retardation}\label{sec:relax_retard_equiv}
To the best of our knowledge, the only theoretical argument for choosing a particular ``branch'' of the general differential constitutive \eqref{eq:differential} is based on Ziegler's thermodynamic orthogonality condition \cite[\S IX-B]{ZiegWeh}, which suggests that one cannot set $\tau_1=\tau_2=\cdots=0$ (``pure retardation'') but must retain some nonzero $\tau_is$. Additionally, it has been shown that ``pure retardation,'' often referred to as Rivlin--Ericksen \cite{RE55} or order $n$ \cite{ColemanNoll_gn}, fluids with only the leading-order terms in the retardation expansion are ill-posed mathematically if one attempts to match the coefficient $\mu_1$ with certain experimental data \citep{FosRaj}.

To better understand the latter result, let us consider a pure relaxation (Maxwell-type) constitutive relation, i.e., keeping only one term on the left-hand side of \eqref{eq:differential}:
\begin{equation}
\left(1+\tau_1 \partial_t \right)\bsf{S} = \mu_0 \bsf{E},\qquad \mu_0,\tau_1 > 0,
\label{eq:const_rel_max}
\end{equation}
which can be rewritten as a pure-retardation constitutive law
\begin{equation}
\bsf{S} = \mu_0 \left(1 + {\tau_1} \partial_t \right)^{-1}\bsf{E} \simeq \mu_0 \left(1-\tau_1 \partial_t \right)\bsf{E}
\label{eq:const_rel_sg}
\end{equation}
for small Deborah numbers, i.e., $\mathrm{De} := \tau_1/t_\mathrm{c} \ll 1$, where $t_\mathrm{c}$ is a characteristic flow time scale (e.g., inverse frequency of oscillation in a rheological experiment \cite[\S3.4]{Bird}).\footnote{The reverse manipulation was used by Cattaneo \cite{Cattaneo} in the derivation of his heat conduction law with finite speed of propagation \cite[p.~376]{JosephPrezAdd}.} Equation \eqref{eq:const_rel_sg} is the constitutive relation for the pure retardation fluid [i.e., keeping only one term on the right-hand side of \eqref{eq:differential}] with $\mu_1 = -\tau_1 < 0$ when $\tau_1 > 0$, which, for unidirectional shear flow, is equivalent to the second-order/grade fluid with the ``bad'' sign (note that $\sgn \mu_1 = -\sgn \Psi_1$) \citep{DunRaj}, as in experiments.

On the other hand, if we were to start with the pure-retardation fluid with the ``good'' sign:
\begin{equation}
\bsf{S} = \mu_0 \left(1 + {\mu_1} \partial_t \right)\bsf{E},\qquad \mu_0,\mu_1>0,
\label{eq:const_rel_sg2}
\end{equation}
then its relaxational ``twin'' has the constitutive relation
\begin{equation}
\left(1 - {\mu_1} \partial_t \right)\bsf{S} \simeq \mu_0 \bsf{E}.
\label{eq:const_rel_max2}
\end{equation}
However, now the coefficient of $\partial_t$ on the left-hand side is negative, giving a noncausal Maxwell model \cite{JP04} with relaxation time $\tau_1 = -\mu_1<0$, which is unphysical. This begs the question: \emph{Can the ``good'' pure-retardation fluid  exist if its pure-relaxation ``twin'' is unphyhsical?}

\section{Well-posedness and Fourier mode analysis} 
The choice of terms in the general differential constitutive relation \eqref{eq:differential} is not entirely arbitrary because the formulation of the viscoelastic memory impacts the resulting model's mathematical well-posedness.

To elucidate this point, consider a one-dimensional shearing motion in the $x$-direction so that the velocity field is $\bm{u} = u(y,t) \hat\imath$. Such a flow linearizes the equations of motions, making it a convenient example. Then, the rate of strain tensor $\bsf{E} \equiv \bm{\nabla}\bm{u} + (\bm{\nabla}\bm{u})^\top$ has only two nonzero components, namely $\mathsf{E}_{xy} = \mathsf{E}_{yx} = {\partial u}/{\partial y}$. Thus, ignoring body forces, the equations of motion for a viscoelastic fluid with a single retardation term (let us call it `RT1') are
\begin{equation}
\varrho_0 \partial_t u = -\partial_x p + \partial_y \mathsf{S}_{yx},\quad
\mathsf{S}_{yx} = \mu_0 \left(1 + \mu_1 \partial_t\right) \mathsf{E}_{xy}.
\label{eq:1D_retard}
\end{equation}

We assume no longitudinal pressure gradient, i.e., $\partial_x p = 0$. Then, eliminating $\mathsf{S}_{xy}$ between the two equations in \eqref{eq:1D_retard}, the evolution equation (see also \cite{ColemanDuffinMizel,T63,P84,CC10,C10}) for the velocity is
\begin{equation}
\varrho_0 \partial_t u - \mu_0\mu_1 \partial_y \partial_t \partial_y u = \mu_0 \partial_y^2 u.
\label{eq:1D_velocity_retard}
\end{equation}
Equation \eqref{eq:1D_velocity_retard} also arises in Euler--Poincar\'e models of ideal fluids with nonlinear dispersion \cite{Holm} and unidirectional flows of the so-called second-order (Rivlin--Ericksen) fluid  \cite[\S6.1]{Bird}.

Now, consider a spatial Fourier mode with wavenumber $k$ and temporal growth rate $\Re\{\sigma\}$: $u(y,t) \propto \Re\{ \re^{\sigma t}\re^{\ri k x} \}$. Substituting the latter into \eqref{eq:1D_velocity_retard} yields the growth rate
\begin{equation}
\sigma_{\mathrm{RT1}}(k) = \frac{-\nu_0 k^2}{ 1 + \mu_1 \nu_0 k^2},
\label{eq:sigma_RT1}
\end{equation}
where $\nu_0 := \mu_0/\varrho_0 (>0)$ is the kinematic viscosity.
If $\Re\{\sigma(k)\} > 0$ for any $k$, then those Fourier modes blow up as $t\to\infty$, which is associated with instability for a linear partial differential equation. Since $\nu_0>0$, the only possibility for instability is if $\mu_1 < 0$, then $\exists k_c := 1/\sqrt{-\mu_1\nu_0}$ such that Fourier modes with $k>k_c$ blowup (short-wave instability).\footnote{In the context of the second-order fluid, various other techniques have also been used to show the intrinsic instability of the RT1 model \eqref{eq:1D_velocity_retard} with $\mu_1<0$  \cite{ColemanDuffinMizel,DunnFos,FosRaj}. At the same time, the experimental data can be fit to the second-order fluid adequately only if first normal stress difference $\Psi_1>0$, which gives $\mu_1 <0$ \cite{MarkovitzBrown,Bird,KhairSquires}, leading to significant controversy in the literature \cite{DunRaj}.}

 
Therefore, the RT1 fluid model is well-posed only if $\mu_1 > 0$. However, as we saw in Section~\ref{sec:relax_retard_equiv}, the RT1 fluid's pure-relaxation twin is noncausal for $\mu_1 > 0$. Could going to the next order in the pure-retardation expansion mitigate the undesirable effects of $\mu_1<0$? The constitutive relation of this (let us call it `RT2') fluid is
\begin{equation}
\mathsf{S}_{yx} = \mu_0\left(1 + \mu_1 \partial_t + \mu_2 \partial_t^2 \right)\mathsf{E}_{yx},
\end{equation}
the evolution equation for its velocity is
\begin{equation}
\varrho_0 \partial_t u - \mu_0\mu_1  \partial_y\partial_t\partial_y u - \mu_0\mu_2 \partial_y\partial_t^2\partial_y u = \mu_0 \partial_y^2 u,
\label{eq:1D_velocity_retard2}
\end{equation}
and the corresponding temporal growth rate has two branches:
\begin{equation}
\sigma_{\mathrm{RT2},\{1,2\}}(k) = \frac{-(1 + \mu_1 \nu_0 k^2) \pm \sqrt{(1 +  \mu_1 \nu_0 k^2)^2 -4 \mu_2 \nu_0^2 k^4}}{2 \mu_2 \nu_0 k^2}.
\end{equation}

We wish to establish whether the second term in the retardation expansion can stabilize the RT1 fluid's instability when $\mu_1 < 0$. To this end, we note that if $\sigma_{\mathrm{RT2},\{1,2\}}\in\mathbb{R}$, then $\sgn \sigma_{\mathrm{RT2},\{1,2\}} = -\sgn (1 + \mu_1 \nu_0 k^2)$, hence 
$\sigma_{\mathrm{RT2},\{1,2\}} > 0$ (blowup) if $k > k_c := 1/\sqrt{-\mu_1\nu_0}$. On the other hand, if $k > k^* = (2\sqrt{\mu_2}\nu_0 - \mu_1\nu_0)^{-1/2}$, where it is clear that $k^* < k_c$, then $\sigma_{\mathrm{RT2},\{1,2\}}\in\mathbb{C}$, and $\Re\{\sigma_{\mathrm{RT2},1}(k)\} = \Re\{\sigma_{\mathrm{RT2},2}(k)\} = -(1 + \mu_1 \nu_0 k^2)/(2 \mu_2 \nu_0 k^2)$. Once again, $\Re\{\sigma_{\mathrm{RT2},\{1,2\}}\} > 0$ (blowup) if $k > k_c$, which is precisely the short-wave instability exhibited by the RT1 fluid!

\begin{figure}[h]
\centerline{\includegraphics[width=0.9\columnwidth]{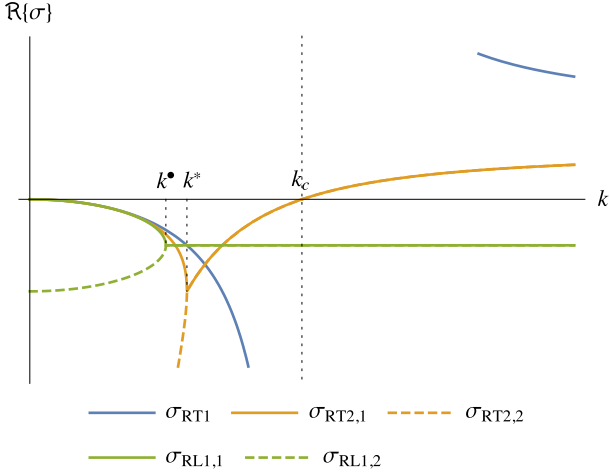}}
\caption{(Color online.) Qualitative illustration of the dependence of the temporal growth rate $\Re\{\sigma\}$ on the wavenumber $k$ of a Fourier mode under the three fluid models considered (with $\mu_1<0$). This plot illustrates coalescence of real parts (i.e., $\sigma\in\mathbb{C}$), for the RT2 and RL1 fluids at $k=k^*$ and $k=k^\bullet$, respectively, and the change of sign of $\Re\{\sigma\}$ at $k=k_c$ for the RT1 and RT2 fluids, which signifies blowup of Fourier modes with $k>k_c$.}
\label{fig:dispersion}
\end{figure}

Clearly, if we require $\mu_1 < 0$ [so that the pure-relaxation twin model \eqref{eq:const_rel_max2} is causal], then the pure-retardation fluids (RT1 and RT2) cannot be ``salvaged'' as mathematical models. As in Section~\ref{sec:relax_retard_equiv}, for $\tilde{\mathrm{De}}:=|\mu_1|/t_\mathrm{c}\ll1$, we can rewrite \eqref{eq:1D_retard}$_2$ as  pure-relaxation (Maxwell-type) model, $\left(1 - \mu_1 \partial_t\right) \mathsf{S}_{yx}  = \mu_0 \mathsf{E}_{yx}$ (let us call it `RL1'), then the equation for the evolution of its velocity (see also \cite{DP71,JP05,JPB04}) is
\begin{equation}
- \mu_1\varrho_0 \partial_t^2 u + \varrho_0 \partial_t u = \mu_0 \partial_y^2 u, \qquad \mu_1 < 0. 
\label{eq:1D_velocity_relax}
\end{equation}
This asymptotically-equivalent Maxwell-type model\footnote{In contrast to Footnote 7, the RL1 model \eqref{eq:1D_velocity_relax} with $\mu_1<0$ has been shown to exhibit continuous dependences on the relaxation time $\tau_1=-\mu_1$, and its solutions converge to those of Navier--Stokes fluid as $\tau_1\to0^+$ \cite{PS99}.} (with $\tau_1 = -\mu_1 > 0$) yields Fourier modes with temporal growth rates
\begin{equation}
\sigma_{\mathrm{RL1},\{1,2\}}(k) = \frac{1 \mp \sqrt{1 + 4\mu_1 \nu_0 k^2} }{2\mu_1},\qquad \mu_1 < 0.
\label{eq:sigma_RL1}
\end{equation}
Clearly, if $\sigma_{\mathrm{RL1},\{1,2\}} \in \mathbb{R}$, then $\sigma_{\mathrm{RL1},\{1,2\}} < 0$ $\forall k$ as long as $\mu_1 <0$ (the causal RL1 case or, equivalently, the ``bad'' RT1 fluid). The two real roots $\sigma_{\mathrm{RL1},\{1,2\}}$ merge at $k = k^\bullet := \left(2\sqrt{|\mu_1|\nu_0}\right)^{-1}$, and $\sigma_{\mathrm{RL},\{1,2\}} \in \mathbb{C}$ for $k>k^\bullet$. Nevertheless, $\Re\{\sigma_{\mathrm{RL1},\{1,2\}}\} = -|2\mu_1|^{-1} < 0$ for $k>k^\bullet$, hence these oscillatory modes decay.

The latter conclusion begs the question: \emph{Could experiments that fit data to a model with a single retardation time (e.g., experiments with flows that linearize the second-grade fluid's equation of motion) actually be predicting $\mu_1 <0$ because the data should, in fact, be fit to a Maxwell-type model with a single relaxation time $\tau_1 = -\mu_1$?}

To the best of our knowledge, this question has not been asked or answered in the literature. Therefore, this brief preliminary research report could lead to a new approach to understanding the difficulties of interpreting experimental measurements of what are assumed to be second-order/grade fluids.

\section{Conclusion} We have suggested that it might be difficult to experimentally distinguish between rheological formulations involving relaxation and retardation.
Upon further research, it is conceivable that this observation could mean that one \emph{cannot} select a pure-retardation differential rheological model [i.e., the ``right branch'' of \eqref{eq:differential}]. We have informally screened a number of experimental papers on high-frequency oscillatory motions of a fluid in a gap (a standard rheological experiment \cite[\S3.4]{Bird}), and we found that silicon oils are very well approximated by a Maxwell-type relaxational law. In these experiments, any effect of a non-zero retardation time could only appear at very high frequencies, beyond the measured ones.\footnote{Notice also that the Taylor series expansions of $\sigma_{\mathrm{RT1}}(k)$, $\sigma_{\mathrm{RT2,1}}(k)$ and $\sigma_{\mathrm{RL1,1}}(k)$, for $|\mu_1|\nu_0 k^2\ll 1$, first differ in the coefficient of $k^6$, meaning all these models can only be distinguished if very high wavenumbers are excited in the experiment.} Hence, a constitutive relation with one term in the relaxation expansion and one term in the retardation expansion:\footnote{Attributed to Sir Harold Jeffreys \cite{Joseph_History} and first studied in detail by Frohlich and Sack \cite{Frohlich}, the model \eqref{eq:1D_relax_retard} is commonly known as the ``Oldroyd-B fluid,'' though Oldroyd's contribution was to make the model frame-indifferent \cite{Oldroyd}, a modification that is not manifested for unidirectional flows in steady domains.}
\begin{equation}
\left(1 + \tau_1 \partial_t\right) \bsf{S}  = \mu_0\left( 1 + \mu_1\partial_t\right)\bsf{E},\qquad \tau_1 > \mu_1 > 0,
\label{eq:1D_relax_retard}
\end{equation}
which has the velocity evolution equation (see also \cite{T62a,C10})
\begin{equation}
\tau_1 \partial_t^2 u + \partial_t u = \nu_0 \partial_y^2 u + \nu_0\mu_1\partial_y\partial_t\partial_y u,
\label{eq:1D_velocity_relax_retard}
\end{equation}
might be most appropriate, in practice, because it incorporates both types of memory effects. Justifying the latter assertion is an avenue of future work.

\section*{Acknowledgements}
The authors thank Dr.\ Pedro M.\ Jordan for helpful remarks and discussions. I.C.C.\ thanks Prof.\ Martin Ostoja-Starzewski for drawing his attention to \cite{ZiegWeh}.

\bibliographystyle{elsarticle-num}
\bibliography{SG}
\end{document}